\documentclass[aps,prl,twocolumn,superscriptaddress,longbibliography]{revtex4-2}
\usepackage{graphicx}
\usepackage{amssymb,amsmath}
\usepackage{bm,bbold}
\usepackage{dcolumn}
\usepackage{float}
\usepackage[OT1]{fontenc} 
\usepackage{url}
\usepackage{mathrsfs}
\usepackage{slashed,comment}
\usepackage{color}
\usepackage{verbatim}
\usepackage{enumitem}
\usepackage{soul,physics}
\usepackage[driverfallback=dvipdfm]{hyperref}
\hypersetup{pdfpagemode=FullScreen,colorlinks=true,breaklinks,urlcolor=blue,linkcolor=blue,citecolor=blue}

\usepackage{subfigure}
\usepackage{amssymb}
\usepackage{bm}
\usepackage{graphicx}
\usepackage{amsmath,amssymb}

\usepackage{pdfpages} 
\usepackage{pgffor} 

\makeatletter
\AtBeginDocument{\let\LS@rot\@undefined}
\makeatother

\def\supplementfilename{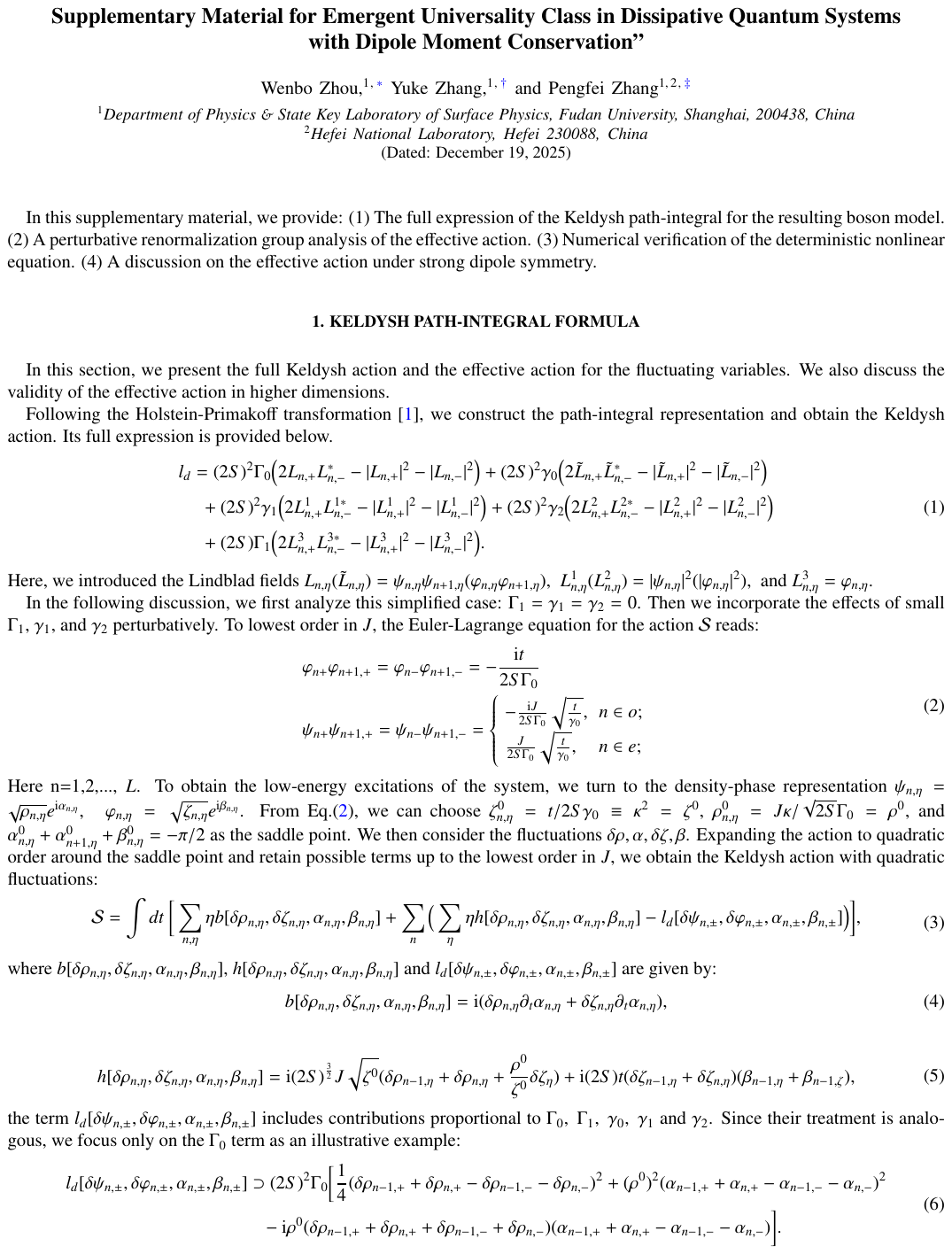}

\pdfximage{\supplementfilename}
\def\numbersupplementpages{\the\pdflastximagepages}

\newif\ifarXiv
\arXivtrue 

\begin{document}
 
  \title{Emergent Universality Class in Dissipative Quantum Systems\\ with Dipole Moment Conservation}

  \author{Wenbo Zhou}
  \thanks{These authors contributed equally to this work.}
  \affiliation{Department of Physics \& State Key Laboratory of Surface Physics, Fudan University, Shanghai, 200438, China}

  \author{Yuke Zhang}
  \thanks{These authors contributed equally to this work.\\yukezhanga08@gmail.com}
  \affiliation{Department of Physics \& State Key Laboratory of Surface Physics, Fudan University, Shanghai, 200438, China}

  \author{Pengfei Zhang}
  \thanks{PengfeiZhang.physics@gmail.com}
  \affiliation{Department of Physics \& State Key Laboratory of Surface Physics, Fudan University, Shanghai, 200438, China}
  \affiliation{Hefei National Laboratory, Hefei 230088, China}

  \date{\today}

  \begin{abstract}
  Understanding the non-equilibrium dynamics of quantum many-body systems remains one of the grand challenges of modern physics. In particular, increasing attention has been devoted to the emergence of non-equilibrium universality classes that have no equilibrium counterparts. A prominent example is the Kardar-Parisi-Zhang universality class realized in dissipative Bose-Einstein condensates. In this Letter, motivated by recent experimental advances, we investigate the universal dynamics of dissipative quantum systems with dipole moment conservation. We develop an effective field theory description, supported by a concrete quantum spin model, to capture the resulting universal behaviors. Our analysis unveils a novel strongly interacting non-equilibrium fixed point that governs the equal-time phase fluctuations in systems with either strong or weak dipole symmetries. Moreover, charge transport becomes subdiffusive in the presence of strong dipole symmetry, while it remains diffusive in the weakly symmetric case. Our results reveal the intricate interplay between kinetic constraints and dissipation in quantum many-body systems.
  \end{abstract}
   
  \maketitle

  \emph{ \color{blue}Introduction.--} Unlike equilibrium systems, whose behaviors are governed by well-established thermodynamic principles, the non-equilibrium dynamics of quantum many-body systems display rich and complex phenomena and remain an active area of research. Of particular interest is the emergence of novel non-equilibrium universality classes in dissipative quantum systems~\cite{RevModPhys.97.025004,2021PhRvB.103d5302M, PhysRevLett.110.195301,PhysRevLett.116.070407,Cheung:2017bhb,Alberton:2020lfq, Xiao:2024arb,Zou:2023rmw}, which govern long-distance behaviors while being largely insensitive to microscopic details. These universality classes are fundamental to understand a wide range of physical systems, from cold atomic gases~\cite{RevModPhys.85.553, Daley:2014fha, PhysRevLett.113.210401, Morsch:2018zwc,Muller:2012bxu,Diehl:2008amb,Zhao:2023xuh,Huh:2023xso,PhysRevA.82.033603,Erne:2018gmz,PhysRevLett.124.040403,PhysRevLett.119.250404,Gazo:2023exc} to condensed matter materials~\cite{RevModPhys.85.299, 2013CMPh...1623008C,   2025NatPh..21.1374M,2025JPhA...58L5004L,2025PhRvR...7c3218B,RevModPhys.82.1489,2014NatPh..10..803B,Li:2024pta,2025PhRvB.112q4313N,RevModPhys.76.663}. For example, in the absence of dissipation, low-energy fluctuations in Bose-Einstein condensates are mediated by phonons with linear dispersion~\cite{PhysRevLett.83.2876,Zhai_2021}. In contrast, dissipation drives the system into a completely different universality class~\cite{2016RPPh...79i6001S,PhysRevLett.110.195301,2014PhRvB..89m4310S,PhysRevX.5.011017} described by the celebrated Kardar-Parisi-Zhang (KPZ) equation ~\cite{PhysRevLett.56.889,PhysRevLett.104.230602,PhysRevLett.78.274}. In one dimension, this leads to a dynamical exponent $z=3/2$, in agreement with numerical simulations~\cite{PhysRevB.92.155307}. Moreover, non-equilibrium universality has also been identified in the heating dynamics of Luttinger liquids~\cite{2015PhRvA..92a3603B,PhysRevB.107.045110,PhysRevB.105.205125}. 
  
  \begin{figure}[t]
    \centering
    \includegraphics[width=0.95\linewidth]{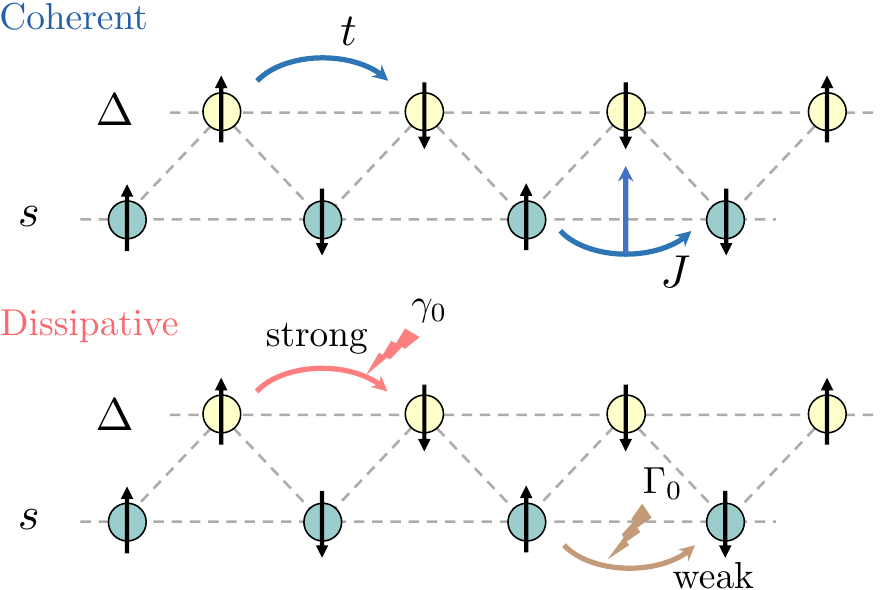}
    \caption{A schematic of our one-dimensional quantum spin model with dipole moment conservation for $S=1/2$. The system consists of two types of spins, denoted by $s$ and $\Delta$. During the coherent evolution governed by the Hamiltonian, exchanges of $s$ spins are accompanied by flips of the $\Delta$ spins, ensuring the dipole moment conservation. For the dissipative dynamics, we consider several processes that are compatible with this symmetry. We provide examples of dominant dissipation processes that exhibit strong or weak dipole symmetry. }
    \label{fig:schemticas}
  \end{figure}

  Meanwhile, the rapid development of highly controllable quantum simulation platforms has enabled experimental studies of quantum systems with kinetic constraints. Recent experiments with cold atoms in optical lattices, for instance, have realized settings in which dipole moments are approximately conserved due to a strong potential gradient~\cite{Kim:2025gyd,PhysRevX.10.011042}. Similar mobility constraints have also been investigated in the context of fracton topological order~\cite{Gromov:2017vir,Ebisu:2024cke,PhysRevB.97.085116,Griffin:2014bta,PhysRevB.98.115134,PhysRevX.9.031035,PhysRevB.98.035111,Gorantla:2022eem,Jain:2021ibh,Gromov:2020yoc}. These constrained dynamics give rise to a variety of novel phenomena, including Stark many-body localization~\cite{PhysRevLett.122.040606,PhysRevB.105.L140201,Morong:2021qyr}, slow charge or operator dynamics ~\cite{PhysRevResearch.2.033129,Cheng:2024fwk,Feldmeier:2021fxa,Khemani:2017nda}, and multi-particle quantum walks~\cite{Kim:2025gyd} . For example, the dipolar Bose-Hubbard model exhibits a quadratic phonon mode~\cite{Lake:2022ico,Lake_2023}, in sharp contrast to the superfluid phase of the conventional Bose-Hubbard model~\cite{Zhai_2021,Sachdev_2011}. Nevertheless, these studies have primarily focused on closed quantum systems, while non-equilibrium dissipative dynamics, which can exhibit fundamentally different universal behaviors, remain largely unexplored.

  In this Letter, inspired by these developments, we study dissipative quantum systems with dipole moment conservation, focusing on the emergence of a non-equilibrium fixed point. We consider concrete quantum spin models (as illustrated in FIG. \ref{fig:schemticas}) and perform a standard large-$S$ expansion to develop an effective field theory description, which is expected to capture universal behavior. We fix the charge conservation to be strongly symmetric~\cite{Bu_a_2012,Albert_2014,albert2018lindbladiansmultiplesteadystates,Lieu_2020} and consider two different scenarios in which dipole conservation is either weakly or strongly symmetric. The results show that the equal-time phase fluctuations in both cases are governed by a novel non-equilibrium fixed point that is intrinsically strongly interacting. In addition, charge transport becomes sub-diffusive in the presence of strong dipole symmetry, while it remains diffusive when dipole conservation is weak. Our findings reveal a rich non-equilibrium universality class in dissipative systems with kinetic constraints and could be relevant for experiments on quantum simulation platforms. 

  \emph{ \color{blue}Model.--} To motivate the effective field-theory description, we first consider a concrete one-dimensional spin model that exhibits weak dipole moment conservation. The modifications required for the strongly symmetric case will be discussed later. The model contains two species of quantum spins with spin-$S$. Their spin operators are denoted by $\hat{s}^\alpha_n$ and $\hat{\Delta}^\alpha_n$, where $\alpha\in \{x,y,z\}$ and $n=1,2,...,L$ labels the lattice sites. The intrinsic dynamics of the system is described by the Hamiltonian
  \begin{equation}\label{eqn:defH}
  \hat{H}=J\sum_{n}\hat{s}^+_{n+1}\hat{s}^-_{n}\hat{\Delta}_{n}^-+t\sum_{n}\hat{\Delta}_{n+1}^+\hat{\Delta}_{n}^-+\text{H.c.},
  \end{equation}
  where we introduce $\hat{s}_n^\pm=\hat{s}_n^x\pm i\hat{s}_n^y$, and define $\hat{\Delta}_n^\pm$ analogously. The system is invariant under global spin rotations of the $s$ spins along the $z$-axis, corresponding to the conservation of $\hat{Q}=\sum_n \hat{s}^z_n$. We refer to $\hat Q$ as the charge of the system. In addition, under open boundary conditions the quantity $\hat{D}=\sum_n (n \hat{s}^z_n+\hat{\Delta}^z_n)$ is conserved, which we refer to as the dipole moment conservation. While we focus on the one-dimensional case, generalizations to higher dimensions are straightforward, with $s$ spins residing on sites and $\Delta$ spins defined on the links. From a symmetry perspective, this spin system belongs to the same universality class as the dipolar Bose-Hubbard model ~\cite{Lake:2022ico,Lake_2023}, which has been realized in a recent experiment ~\cite{Kim:2025gyd}. Similar correlated hopping terms have also been studied both theoretically and experimentally in the context of quantum simulations of lattice gauge theories ~\cite{Cheng:2024nep, PhysRevX.10.021041,PRXQuantum.3.040317,Halimeh:2023lid,Yang:2020yer,Zhou:2021kdl,Wang:2022dpp,2005PhRvL..95d0402B,RevModPhys.51.659,Chandrasekharan:1996ih}.

   Next, we incorporate the effects of dissipation by coupling the system to a Markovian bath. The evolution of the density matrix is then governed by the Lindblad master equation~\cite{1976JMP....17..821G, 1976CMaPh..48..119L} $ \partial_t\rho=-\mathrm{i}[\hat{H},\rho]+\mathcal{L}_d[\rho]$, with the dissipative superoperator
   \begin{equation}\label{eqn:defL}
   \begin{aligned}
   \mathcal{L}_d=&\sum_{\langle mn \rangle}\Gamma_0\mathcal{D}[\hat{s}^-_{m\in o}\hat{s}_{n\in e}^+]+\gamma_0\mathcal{D}[\hat{\Delta}^-_{m\in o}\hat{\Delta}_{n\in e}^{+}]\\
   &+\sum_m\Gamma_1\mathcal{D}[\hat{\Delta}^-_{m}]+\gamma_1\mathcal{D}[\hat{s}^z_{m}]+\gamma_2\mathcal{D}[\hat{\Delta}^z_{m}].
   \end{aligned}
   \end{equation}
   Here, we divide the lattice into even ($e$) and odd ($o$) sites and define the dissipator for a Lindblad operator $\hat{L}$ as $\mathcal{D}[\hat{L}](\rho)=2\hat{L}\rho \hat{L}^{\dagger}-\{\hat{L}^{\dagger}\hat{L},\rho\}$. All Lindblad operators in Eq.~\eqref{eqn:defL} commute with the charge operator $Q$, and therefore no charge is exchanged between the system and the bath. In the recent terminology, this corresponds to a strong symmetry associated with the charge~\cite{Bu_a_2012,Albert_2014,albert2018lindbladiansmultiplesteadystates,Lieu_2020}. In contrast, for the dipole moment we have $[\hat{s}^-_{m\in o}\hat{s}_{n\in e}^+,\hat{D}]=\pm \hat{s}^-_{m\in o}\hat{s}_{n\in e}^+$ for nearest-neighbor sites $(m,n)$, and $[\hat{\Delta}_n^-,\hat{D}]=\hat{\Delta}^-_n$. Thus, there are Lindblad operators that change the dipole moment $D$. Nevertheless, they do not induce any coherence between states with different dipole moments. For example, we have $$\mathcal{D}[\hat{s}^-_{m\in o}\hat{s}_{n\in e}^+](\hat{U}_D\rho \hat{U}_D^\dagger)=\hat{U}_D\mathcal{D}[\hat{s}^-_{m\in o}\hat{s}_{n\in e}^+](\rho )\hat{U}_D^\dagger$$ for the symmetry operation $\hat{U}_D(\beta_0)=e^{-i\beta_0 \hat{D}}$. Therefore, the dissipator, and thus the entire Lindblad master equation, preserves the weak symmetry associated with the dipole moment. Finally, we remark that since our ultimate goal is to construct a universal effective field theory, we allow arbitrarily small perturbations to both the Hamiltonian and the Lindblad terms, provided they are compatible with the relevant symmetries.
   
  \emph{ \color{blue}Keldysh field theory.--} In this work, we focus on the strong-dissipation limit $\Gamma_0, \gamma_0 \gg J, t$ and analyze the transport properties near the steady state. When $J = t = 0$, the dissipator in Eq.~\eqref{eqn:defL} drives the system toward a state in which all spins on odd/even sites are polarized along the $\mp z$ direction:
  \begin{equation}\label{eq:polarized}
  \langle \hat{s}_{n\in o}^z\rangle= \langle \hat{\Delta}_{n\in o}^z\rangle=-S,\ \ \ \ \langle \hat{s}_{n\in e}^z\rangle= \langle \hat{\Delta}_{n\in e}^z\rangle=S.
  \end{equation}  
  Adding small $J$ and $t$ then allows the spins to fluctuate around this polarized state, motivating the application of the Holstein-Primakoff transformation~\cite{Altland:2006si}:
  \begin{equation}
  \begin{aligned}
  \left\{
  \begin{array}
  {ll}\hat{s}_n^z=\hat \psi_n^\dagger \hat \psi_n-S,\quad \hat{s}_n^+=\hat \psi_n^\dagger\sqrt{2S-\hat \psi_n^\dagger \hat \psi_n}, & n\in o, \\
  \hat{s}_n^z=S-\hat \psi_n^\dagger \hat \psi_n,\quad \hat{s}_n^+=\sqrt{2S-\hat \psi_n^\dagger \hat \psi_n}\hat \psi_n, & n\in e.
  \end{array}\right.
  \label{eq:largeS}
  \end{aligned}
  \end{equation}
  Here, $\hat{\psi}_n$ is the bosonic annihilation operator on site $n$, satisfying the canonical commutation relation $[\hat{\psi}_n, \hat{\psi}_m^\dagger] = \delta_{nm}$. A similar definition relates the spin operators $\hat{\Delta}_n^\alpha$ to bosonic operators $\hat{\varphi}_n$. The vacuum state of the bosons is chosen as the polarized state \eqref{eq:polarized}, ensuring a small boson occupation in the strong-dissipation limit. This justifies the large-$S$ expansion in Eq.~\eqref{eq:largeS}, where $\sqrt{2S-\hat \psi_n^\dagger \hat \psi_n}$ is approximated by $\sqrt{2S}$. 

  We apply the Keldysh path-integral approach~\cite{kamenev2023field} to analyze the resulting boson model. The contour consists of a forward evolution branch with bosonic fields $(\psi_{n,+}, \varphi_{n,+})$ and a backward evolution branch with bosonic fields $(\psi_{n,-}, \varphi_{n,-})$. The partition function of the system reads $\mathbb{1}=\int \prod_{\eta,n} D\psi_{n,\eta} D\varphi_{n,\eta}~e^{-\mathcal{S}}$ with $\eta=\pm$ and  
  \begin{equation}
  \begin{aligned}
  \mathcal{S}=&\int dt~\bigg[\sum_{n,\eta} \eta(\bar{\psi}_{n,\eta}\partial_t\psi_{n,\eta}+\bar{\varphi}_{n,\eta}\partial_t\varphi_{n,\eta})\\
  &+\sum_{n}\Big(\sum_{\eta}i\eta h[\psi_{n,\eta},\varphi_{n,\eta}]-l_d[\psi_{n,\pm},\varphi_{n,\pm}]\Big)\bigg],
  \end{aligned}
  \end{equation}
  where the Hamiltonian contribution $h$ is given by
  \begin{equation}
  h[\psi_{n},\varphi_{n}]={(2S)}^\frac{3}{2}J\psi_{n+1}\psi_n\varphi_n+2St\varphi_{n+1}\varphi_n+\text{C.c.}
  \end{equation}
  and the dissipation part $l_d$ takes the form of 
  \begin{equation}
  \begin{aligned}
  l_d=& (2S)^2\Gamma_0 \Big(2L_{n,+}L_{n,-}^{*}-|L_{n,+}|^2-|L_{n,-}|^2\Big)\\
  &+(2S)^2\gamma_0 \Big(2\tilde{L}_{n,+}\tilde{L}_{n,-}^{*}-|\tilde{L}_{n,+}|^2-|\tilde{L}_{n,-}|^2\Big).
  \end{aligned}
  \end{equation}
  Here, we introduced Lindblad fields $L_{n,\eta}=\psi_{n,\eta}\psi_{n+1,\eta}$ and $\tilde{L}_{n,\eta}=\varphi_{n,\eta}\varphi_{n+1,\eta}$. For conciseness, we set $\Gamma_1 = \gamma_1 = \gamma_2 = 0$, and the full expression is provided in the supplementary material for completeness~\cite{SuppMat}. In the following discussion, we first analyze this simplified case and then incorporate the effects of small $\Gamma_1$, $\gamma_1$, and $\gamma_2$ perturbatively. The strong charge symmetry and weak dipole symmetry also manifests in the path-integral representation. The action $\mathcal{S}$ is invariant under the transformation 
  \begin{equation}\label{eq:symmetryKeldysh}
  \begin{aligned}
  &\psi_{n,\eta}\rightarrow e^{i(-1)^n\beta_0 n}e^{i(-1)^n\alpha_{0,\eta}}\psi_{n,\eta},\\ &\varphi_{n,\eta}\rightarrow e^{i(-1)^n\beta_0 }\varphi_{n,\eta}.
  \end{aligned}
  \end{equation}
  where $\alpha_{0,+}$ and $\alpha_{0,-}$ are independent phases originating from charge conservation, and $\beta_0$ represents the phase associated with the symmetry operation $\hat{U}_D(\beta_0)$ generated by the dipole moment $\hat{D}$.

  \emph{ \color{blue}Effective action.--} We are interested in analyzing the modes that dominates the long-wave length fluctuations of the system. The symmetry \eqref{eq:symmetryKeldysh} indicates the relevant modes are the phase fluctuations, which govern the low-energy transport. To extract the effective theory, we first employ the density-phase representation by introducing $$\psi_{n,\eta} = \sqrt{\rho_{n,\eta}}, e^{i\alpha_{n,\eta}},\ \ \ \varphi_{n,\eta} = \sqrt{\zeta_{n,\eta}}, e^{i\beta_{n,\eta}}.$$ 
  Then, we perform a saddle-point analysis with respect to both the density and phase fields. This yields the saddle-point solution as $\zeta_{n,\eta}^0=t/2S\gamma_0\equiv \kappa^2$, $\rho_{n,\eta}^0=J\kappa/\sqrt{2S}\Gamma_0$, and $\alpha_{n,\eta}^0+\alpha_{n+1,\eta}^0+\beta_{n,\eta}^0=-\pi/2$. Note that the saddle-point value is independent of $\eta$, which follows from the symmetry between the forward and backward time branches. 

  Next, we examine the quadratic fluctuations around the saddle point by deriving the effective action. The full derivation is provided in the supplementary material~\cite{SuppMat}. Here, we summarize only the key steps: (i) We expand the fields as $\rho_{n,\eta}=\rho_{n,\eta}^0+\delta \rho_{n,\eta}$, and analogously for the other fields. (ii) Motivated by Eq.~\eqref{eq:symmetryKeldysh}, the continuum fields are defined as
  \begin{equation}
  \delta\rho_\eta(x)\equiv (-1)^n \delta\rho_{n,\eta},\ \ \ \ \ \alpha_\eta (x) \equiv (-1)^n \delta\alpha_{n,\eta}.
  \end{equation}
  and similarly for $\delta\zeta_\eta(x)$ and $\beta_\eta(x)$. (iii) We perform the standard Keldysh rotation to separate classical and quantum components~\cite{kamenev2023field}, $\delta\rho_{c/q}(x)=\frac{1}{\sqrt{2}}(\delta\rho_{+}(x)\pm\delta\rho_{-}(x))$ and analogously for the other fields. (iv) Finally, after integrating out fields except $\alpha_c(x)$, $\alpha_q(x)$, $\delta \rho_c(x)$, and $\delta \rho_q(x)$, we obtain the quadratic effective action $\mathcal{S}_{\text{eff}}=\int dx dt~\mathcal{L}^w_{1}[\delta \rho_c,\alpha_q]+\mathcal{L}_{2}[\delta \rho_q,\alpha_c]$, with 
  \begin{equation}\label{eq:Seff}
  \begin{aligned}
  &\mathcal{L}^w_{1}[\delta \rho_c,\alpha_q]=i\delta \rho_{c}(\partial_t\alpha_q+D\nabla^2\alpha_q)+\tilde{C}(\nabla\alpha_q)^2/2,\\&
  \mathcal{L}_{2}[\delta \rho_q,\alpha_c]=i\delta \rho_q(\partial_t\alpha_c+\tilde D\nabla^4\alpha_c)+C\delta \rho_q^2/2.
  \end{aligned}
  \end{equation}
  Here, coefficients $C,\tilde{C},D,\tilde{D}$ are combinations of microscopic parameters, whose explicit forms are not essential for the prediction of universal features. There is a decoupling between the classical and quantum components of the phase field, consistent with previous studies of models without dipole-moment conservation. In the supplementary material~\cite{SuppMat}, we further argue that the effective action remains valid for an arbitrary spatial dimension $d$.

  The Lagrangian density $\mathcal{L}^w_{1}$ is invariant (up to a total derivative) under a constant shift $\alpha_q(x,t) \rightarrow \alpha_q(x,t) + c_0$, reflecting strong charge conservation symmetry. As elaborated in Ref.~\cite{Gu:2024wgc}, this action gives rise to long-range order $\text{lim}_{r\rightarrow \infty}\langle e^{i\alpha_q(\mathbf{r})}e^{-i\alpha_q(0)}\rangle \neq 0$ in arbitrary dimensions, as well as the charge diffusion $\langle \delta \rho_{c}(\mathbf{r},t)\delta \rho_{c}(0,0)\rangle = \frac{1}{(4\pi Dt)^{d/2}}e^{-\frac{r^2}{4Dt}}$ with dynamical exponent $z=2$~\cite{Ogunnaike_2023,Moudgalya_2024,Huang_2025,Gu:2024wgc}. These results imply that the strong charge conservation symmetry is spontaneously broken to a weak symmetry~\cite{Lee_2023,Lessa_2025,sala2024spontaneousstrongsymmetrybreaking,Huang_2025,Gu:2024wgc,Ziereis:2025rit,sun2025schemedetectstrongtoweaksymmetry,Kuno_2024,zhang2025strongtoweakspontaneousbreaking1form,liu2024diagnosingstrongtoweaksymmetrybreaking,Weinstein_2025,Guo_2025,song2025strongtoweakspontaneoussymmetrybreaking,S__2025}. We highlight that this behavior is qualitatively different from that of a dipole-conserving system without dissipation, which exhibits subdiffusion~\cite{PhysRevResearch.2.033129,Cheng:2024fwk,Feldmeier:2021fxa}. The weak dipole-moment conservation in an open system allows dipole moment to be exchanged between the system and the bath, thereby restoring an ordinary charge-diffusion process. This is also consistent with the observation that the symmetry transformation parametrized by $\beta_0$ in Eq.~\eqref{eq:symmetryKeldysh} does not couple to $\alpha_q$.

  \emph{ \color{blue}Universal phase fluctuations.--} In contrast, the classical phase field transforms as $\alpha_c(x,t)\rightarrow \alpha_c(x,t)+\sqrt{2}\beta_0 x$ under the dipole symmetry operation. Consequently, the Lagrangian density $\mathcal{L}_{2}$ must remain invariant under this transformation, consistent with the explicit form in Eq.~\eqref{eq:Seff}. Naively, $\mathcal{L}_{2}$ predicts subdiffusion with a dynamical exponent $z = 4$. However, the quadratic action Eq.~\eqref{eq:Seff} obtained from the concrete spin model exhibits an enhanced symmetry, permitting the classical phase field to shift by an arbitrary polynomial in position up to third order. This enhanced symmetry can be broken once interactions between the fluctuation fields are included. From the effective field theory perspective, we should incorporate all relevant terms compatible with the symmetry, provided they can arise either from deformations of the original model or through the renormalization-group flow. After including most relevant terms, the effective action becomes
  \begin{equation}\label{eq:Seff_int}
  \tilde{\mathcal{L}}_{2}=i\delta \rho_q(\partial_t\alpha_c-\tilde{D}_2 \nabla^2\alpha_c+\tilde D\nabla^4\alpha_c-g( \nabla^2\alpha_c)^2)+\frac{C}{2}\delta \rho_q^2.
  \end{equation}
  Here, the interaction term $\delta \rho_q (\nabla^{2}\alpha_c)^{2}$ can arise by adding a four-body hopping term ($\hat{s}_{n}^{+}\hat{s}_{n+2}^{-}\hat{s}_{n+4}^{-}\hat{s}_{n+6}^{+}$) to the original Hamiltonian. This term has a structure similar to the interaction term in the field-theoretical representation of the KPZ equation ~\cite{2016RPPh...79i6001S,Gu:2024wgc}. The quadratic term $\delta \rho_q \tilde{D}_2 \nabla^{2}\alpha_c$ then emerges upon evaluating the one-loop correction generated by this interaction, as detailed in the supplementary material~\cite{SuppMat}. 

  \begin{figure}[t]
    \centering
    \includegraphics[width=0.99\linewidth]{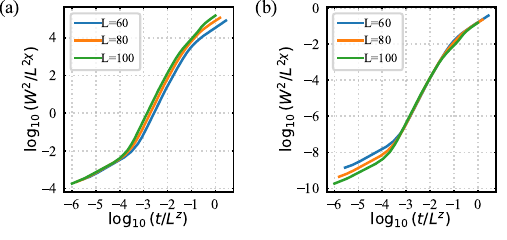}
    \caption{Numerical simulations of the Langevin equation \eqref{eq:classical} with $g = 0.5$, $\tilde{D}_2 =0 $, $\tilde{D} =1 $, and $C = 1$ are performed for different system sizes $L\in\{60,80,100\}$. We plot the results by fixing $z=2$ and using two different exponents: (a) $\chi = 1/2$ and (b) $\chi = 2$. The results clearly demonstrate the emergence of a novel non-equilibrium universality class in the long-time limit.  }
    \label{fig:num}
  \end{figure}

  Without the interaction $g$, the free theory is described by assuming $\tilde{D}_2$ and $C$ are marginal. This gives the scaling dimensions $[\alpha_c] = (d-2)/2$ and $[\delta \rho_q] = (d+2)/2$, as well as the dynamical exponent $z = 2$. The interaction term then has scaling $[g] = -(d+2)/2 < 0$ and is therefore irrelevant. This corresponds to a consistent fixed point, known as the Edwards-Wilkinson (EW) universality class~\cite{10.1098/rspa.1982.0056,PhysRevB.110.014203}, which describes diffusion equations with a random noise term. This result is physically reasonable, even though it is obtained perturbatively. 

  However, the numerical results in one dimension reveal a different scaling behavior. Our numerics is carried out by mapping the effective action \eqref{eq:Seff_int} to a Langevin equation for $\alpha_c$ through a Hubbard-Stratonovich transformation of $\delta \rho_q^2$. The resulting equation reads 
  \begin{equation}\label{eq:classical}
  \partial_t \alpha_c=\tilde{D}_2 \nabla^2\alpha_c-\tilde D\nabla^4\alpha_c+g( \nabla^2\alpha_c)^2+\xi,
  \end{equation}
  where $\xi(x,t)$ is Gaussian white noise satisfying $\langle \xi(x,t) \rangle=0$ and $\langle \xi(x,t) \xi(x',t’)\rangle=C\delta(x-x')\delta(t-t')$. In the numerical simulations, we initialize the system with $\alpha_c(x,0) = 0$ and perform the evolution by discretizing both space and time. Analogous to previous studies of the KPZ universality class \cite{PhysRevLett.56.889,PhysRevLett.104.230602,PhysRevLett.78.274}, we compute the roughness function~\cite{PhysRevLett.56.889,1985JPhA...18L..75F}: 
  \begin{equation}
  \begin{aligned}
  \left(W(t,L)\right)^2\equiv\frac{1}{L}\int_0^L\left(\alpha_c(x,t)-\overline{\alpha}_c(t)\right)^2\mathrm{d}x,
  \label{eq:def2}
  \end{aligned}
  \end{equation} 
  where $\overline{\alpha}_c(t)$ denotes the spatial average. In the long-time regime, we expect this quantity to exhibit the scaling form $W(t,L) \sim L^{\chi} F(t/L^{z})$, where the exponent $\chi = -[\alpha_c]$ follows from dimensional analysis. The EW universality class then predicts $\chi=1/2$ in one dimension. In FIG.~\ref{fig:num}(a), we present the numerical results for $g=0.5$, $\tilde{D}_2=0$, $\tilde{D}=1$, and $C=1$ using $\chi=1/2$. It clearly shows a growing deviation as time extends beyond the early-time regime. In contrast, the results at long times exhibit a clear data collapse if we instead use $\chi=2$. This signifies the emergence of a new non-equilibrium fixed point that are intrinsically interacting.

  To understand this new fixed point, an important observation is that when the interaction term $g \neq 0$, the coupling constant $\tilde{D}_2$ can be freely shifted as:
  \begin{equation}
  \alpha_c(x,t)\rightarrow \alpha_c(x,t)+\frac{c_0}{2} x^2+ gc_0^2t,\ \ \ \ \tilde{D}_2 \rightarrow \tilde{D}_2+2gc_0.
  \end{equation}
   In particular, we can choose $c_0$ such that $\tilde{D}_2 + 2g c_0$ vanishes, thereby eliminating the quadratic term. This redefinition of $\alpha_c$ also leaves the roughness function $W(t,L)$ invariant. This transformation implies that the terms $\delta \rho_q \nabla^2 \alpha_c$ and $\delta \rho_q (\nabla^2 \alpha_c)^2$ should be simultaneously marginal near the fixed point, which immediately leads to $[\alpha_c] = -2$ and hence $\chi = 2$ for arbitrary spatial dimension $d$. We can then determine $[\delta \rho_q] = d + 2$. Consequently, the quadratic term $\delta \rho_q^2$ is irrelevant and can be neglected in the long-time limit. This reduces the  stochastic equation to a deterministic nonlinear equation:
   \begin{equation}
   \partial_t \alpha_c=-\tilde D\nabla^4\alpha_c+g( \nabla^2\alpha_c)^2.
   \end{equation}
   In the supplementary material, we provide additional numerical demonstration that the random Gaussian noise $\xi(x,t)$ is irrelevant in the long-time limit~\cite{SuppMat}. We emphasize that the deterministic nature ensures that the naive power counting yields the exact scaling dimensions, as no quantum fluctuations are present. Moreover, adding arbitrary higher-order marginal terms, such as $\delta \rho_q (\nabla^2 \alpha_c)^3$, does not modify any scaling dimension.

   The effective action governs the long-distance correlations of the boson fields $\psi_n$. We can estimate the equal-time correlator as $\langle \psi_{n} \overline{\psi}_{0} \rangle \propto \langle e^{i\alpha_c(x)}e^{-i\alpha_c(0)}\rangle  \sim  \exp(-c |x|^{2\chi})$ for some constant $c$ ~\cite{2016RPPh...79i6001S}. This scaling behavior, dictated by our newly identified fixed point, indicates the absence of long-range or even quasi–long-range order. As a result, the weak symmetry associated with charge conservation remains unbroken in any spatial dimension, in contrast to the dissipationless case ~\cite{Lake:2022ico,Lake_2023}. In terms of the original spin model, the correlator of the bosonic fields directly maps to the spin-spin correlation function $\langle \hat{s}_n^+\hat{s}_0^- \rangle$, which is, in principle, experimentally accessible.

   \emph{ \color{blue}Strong dipole symmetry.--} Now, we discuss the modification of the effective field theory when the dipole conservation becomes a strong symmetry. This requires replacing the terms with dissipation part \eqref{eqn:defL} with
  \begin{equation}\label{eqn:defL2}
   \begin{aligned}
   \mathcal{L}_d'&=\sum_{l}\tilde{\gamma}_0\Big(\mathcal{D}[\hat{\Delta}^-_{2l-1}\hat{s}^-_{2l-1}\hat{s}_{2l}^+]+\mathcal{D}[\hat{\Delta}^+_{2l}\hat{s}^-_{2l+1}\hat{s}_{2l}^+]\Big)\\
   &+\sum_{\langle mn \rangle}\gamma_0\mathcal{D}[\hat{\Delta}^-_{m\in o}\hat{\Delta}_{n\in e}^{+}]+\sum_m\gamma_1\mathcal{D}[\hat{s}^z_{m}]+\gamma_2\mathcal{D}[\hat{\Delta}^z_{m}].
   \end{aligned}
   \end{equation}
   Each Lindblad operator in Eq.~\eqref{eqn:defL2} commutes with the dipole moment $\hat{D}$, rendering the strong dipole symmetry. Following a procedure similar to the previous discussions, we can map the system to a Keldysh field theory of bosonic fields, whose action must respect the symmetry:
   \begin{equation}\label{eq:symmetryKeldysh2}
  \begin{aligned}
  &\psi_{n,\eta}\rightarrow e^{i(-1)^n\beta_{0,\eta} n}e^{i(-1)^n\alpha_{0,\eta}}\psi_{n,\eta},\\ &\varphi_{n,\eta}\rightarrow e^{i(-1)^n\beta_{0,\eta} }\varphi_{n,\eta}.
  \end{aligned}
  \end{equation}
  Here, $\beta_{0,+}$ and $\beta_{0,-}$ are independent parameters originating from the strong dipole conservation. Consequently, both the classical phase field $\alpha_c(x,t)$ and the quantum phase field $\alpha_q(x,t)$ couple to the dipole symmetry and are subject to corresponding constraints in the effective action. Since these constraints have already been incorporated into $\mathcal{L}_{2}[\delta \rho_q,\alpha_c]$, it is expected to remain valid in the strongly symmetric case, and all discussions on phase fluctuations in the previous section continue to apply. On the contrary, $\mathcal{L}^w_{1}[\delta \rho_c,\alpha_q]$ should be modified such that $\alpha_q(x,t)\rightarrow \alpha_q(x,t)+c_0 x$ becomes a symmetry. In the supplementary material~\cite{SuppMat}, we show that the resulting action can be obtained by replacing $\nabla \rightarrow \nabla^2$, which leads to the modified effective action:
  \begin{equation}
  \mathcal{L}_{1}^s[\delta \rho_c,\alpha_q]=i\delta \rho_{c}(\partial_t\alpha_q-D\nabla^4\alpha_q)+\tilde{C}(\nabla^2\alpha_q)^2/2.
  \end{equation}
  It is then straightforward to determine the correlation functions. Similar to the weakly symmetry case, the quantum phase field exhibits long-range correlations $\text{lim}_{r\rightarrow \infty}\langle e^{i\alpha_q(\mathbf{r})}e^{-i\alpha_q(0)}\rangle \neq 0$. This is consistent with the general argument for the emergence of strong-to-weak symmetry breaking in Lindbladian systems~\cite{Ziereis:2025rit}. The density correlator takes the form $\langle \delta \rho_c(\mathbf{k},t)\delta \rho_c(-\mathbf{k},0)\rangle\propto e^{-Dtk^4}$, which implies a return probability $\langle \delta \rho_c(\mathbf{r},t)\delta \rho_c(\mathbf{r},0) \rangle \sim t^{-d/4}$. This corresponds to the expected subdiffusive behavior, identical to that of dipole-conserving systems without dissipation, as observed experimentally in Ref.~\cite{PhysRevX.10.011042}.

   \emph{ \color{blue}Discussions.--} In this work, we investigate dissipative quantum systems with dipole-moment conservation using an effective field-theory framework supported by concrete quantum spin models. We assume strong charge conservation while allowing the dipole symmetry to be either weak or strong. In both scenarios, the equal-time phase fluctuations are governed by a novel, intrinsically interacting emergent universality class that exhibits qualitative distinctions from its dissipationless counterpart. Moreover, charge transport remains diffusive when the dipole symmetry is weak but becomes subdiffusive under strong dipole conservation. These results offer a concrete example of the rich interplay between kinetic constraints and dissipation in quantum many-body systems.

   We conclude our study with several remarks. First, although our focus is on open quantum systems, the resulting effective theory may also arise in classical non-equilibrium settings, potentially in surface-growth problems with kinetic constraints~\cite{2020NatSR..1013057A,PhysRevLett.56.889}. Second, while we concentrate on dipole-moment conservation, the extension to higher multipole conservation is straightforward and may lead to additional emergent universality classes~\cite{AnomalousDiffusioninDipole-andHigher-Moment-ConservingSystems}. It is also intriguing to explore the effects of long-range couplings~\cite{Cheng:2024fwk}. Finally, developing algorithms capable of large-scale numerical simulations of our model would be particularly valuable for providing direct verification of our predictions. We leave these directions to future work.
   
\vspace{5pt}
\textit{Acknowledgement.} 
This project is supported by the NSFC under grant 12374477, the Shanghai Rising-Star Program under grant number 24QA2700300, and the Innovation Program for Quantum Science and Technology 2024ZD0300101.

  \bibliography{main.bbl}

\ifarXiv


\section*{Supplementary material}

# Supplementary Material for Emergent Universality Class in Dissipative Quantum Systems with Dipole Moment Conservation"

Wenbo Zhou,[1, *] Yuke Zhang,[1, †] and Pengfei Zhang[1, 2, ‡]

[1]*Department of Physics & State Key Laboratory of Surface Physics, Fudan University, Shanghai, 200438, China*
[2]*Hefei National Laboratory, Hefei 230088, China*
(Dated: December 19, 2025)

In this supplementary material, we provide: (1) The full expression of the Keldysh path-integral for the resulting boson model. (2) A perturbative renormalization group analysis of the effective action. (3) Numerical verification of the deterministic nonlinear equation. (4) A discussion on the effective action under strong dipole symmetry.

## 1. KELDYSH PATH-INTEGRAL FORMULA

In this section, we present the full Keldysh action and the effective action for the fluctuating variables. We also discuss the validity of the effective action in higher dimensions.

Following the Holstein-Primakoff transformation [1], we construct the path-integral representation and obtain the Keldysh action. Its full expression is provided below.

$$\begin{aligned}
l_d = &(2S)^2\Gamma_0\big(2L_{n,+}L_{n,-}^* - |L_{n,+}|^2 - |L_{n,-}|^2\big) + (2S)^2\gamma_0\big(2\tilde{L}_{n,+}\tilde{L}_{n,-}^* - |\tilde{L}_{n,+}|^2 - |\tilde{L}_{n,-}|^2\big) \\
&+ (2S)^2\gamma_1\big(2L_{n,+}^1 L_{n,-}^{1*} - |L_{n,+}^1|^2 - |L_{n,-}^1|^2\big) + (2S)^2\gamma_2\big(2L_{n,+}^2 L_{n,-}^{2*} - |L_{n,+}^2|^2 - |L_{n,-}^2|^2\big) \\
&+ (2S)\Gamma_1\big(2L_{n,+}^3 L_{n,-}^{3*} - |L_{n,+}^3|^2 - |L_{n,-}^3|^2\big).
\end{aligned} \quad (1)$$

Here, we introduced the Lindblad fields $L_{n,\eta}(\tilde{L}_{n,\eta}) = \psi_{n,\eta}\psi_{n+1,\eta}(\varphi_{n,\eta}\varphi_{n+1,\eta})$, $L_{n,\eta}^1(L_{n,\eta}^2) = |\psi_{n,\eta}|^2(|\varphi_{n,\eta}|^2)$, and $L_{n,\eta}^3 = \varphi_{n,\eta}$.

In the following discussion, we first analyze this simplified case: $\Gamma_1 = \gamma_1 = \gamma_2 = 0$. Then we incorporate the effects of small $\Gamma_1$, $\gamma_1$, and $\gamma_2$ perturbatively. To lowest order in $J$, the Euler-Lagrange equation for the action $\mathcal{S}$ reads:

$$\begin{aligned}
\varphi_{n+}\varphi_{n+1,+} = \varphi_{n-}\varphi_{n+1,-} &= -\frac{it}{2S\Gamma_0} \\
\psi_{n+}\psi_{n+1,+} = \psi_{n-}\psi_{n+1,-} &= \begin{cases} -\frac{iJ}{2S\Gamma_0}\sqrt{\frac{t}{\gamma_0}}, & n \in o; \\ \frac{J}{2S\Gamma_0}\sqrt{\frac{t}{\gamma_0}}, & n \in e; \end{cases}
\end{aligned} \quad (2)$$

Here n=1,2,..., $L$. To obtain the low-energy excitations of the system, we turn to the density-phase representation $\psi_{n,\eta} = \sqrt{\rho_{n,\eta}}e^{i\alpha_{n,\eta}}$, $\varphi_{n,\eta} = \sqrt{\zeta_{n,\eta}}e^{i\beta_{n,\eta}}$. From Eq.(2), we can choose $\zeta_{n,\eta}^0 = t/2S\gamma_0 \equiv \kappa^2 = \zeta^0$, $\rho_{n,\eta}^0 = J\kappa/\sqrt{2S}\Gamma_0 = \rho^0$, and $\alpha_{n,\eta}^0 + \alpha_{n+1,\eta}^0 + \beta_{n,\eta}^0 = -\pi/2$ as the saddle point. We then consider the fluctuations $\delta\rho, \alpha, \delta\zeta, \beta$. Expanding the action to quadratic order around the saddle point and retain possible terms up to the lowest order in $J$, we obtain the Keldysh action with quadratic fluctuations:

$$\mathcal{S} = \int dt\Big[\sum_{n,\eta}\eta b[\delta\rho_{n,\eta}, \delta\zeta_{n,\eta}, \alpha_{n,\eta}, \beta_{n,\eta}] + \sum_{n}\Big(\sum_{\eta}\eta h[\delta\rho_{n,\eta}, \delta\zeta_{n,\eta}, \alpha_{n,\eta}, \beta_{n,\eta}] - l_d[\delta\psi_{n,\pm}, \delta\varphi_{n,\pm}, \alpha_{n,\pm}, \beta_{n,\pm}]\Big)\Big], \quad (3)$$

where $b[\delta\rho_{n,\eta}, \delta\zeta_{n,\eta}, \alpha_{n,\eta}, \beta_{n,\eta}]$, $h[\delta\rho_{n,\eta}, \delta\zeta_{n,\eta}, \alpha_{n,\eta}, \beta_{n,\eta}]$ and $l_d[\delta\psi_{n,\pm}, \delta\varphi_{n,\pm}, \alpha_{n,\pm}, \beta_{n,\pm}]$ are given by:

$$b[\delta\rho_{n,\eta}, \delta\zeta_{n,\eta}, \alpha_{n,\eta}, \beta_{n,\eta}] = i(\delta\rho_{n,\eta}\partial_t\alpha_{n,\eta} + \delta\zeta_{n,\eta}\partial_t\alpha_{n,\eta}), \quad (4)$$

$$h[\delta\rho_{n,\eta}, \delta\zeta_{n,\eta}, \alpha_{n,\eta}, \beta_{n,\eta}] = i(2S)^{\frac{3}{2}}J\sqrt{\zeta^0}(\delta\rho_{n-1,\eta} + \delta\rho_{n,\eta} + \frac{\rho^0}{\zeta^0}\delta\zeta_\eta) + i(2S)t(\delta\zeta_{n-1,\eta} + \delta\zeta_{n,\eta})(\beta_{n-1,\eta} + \beta_{n-1,\zeta}), \quad (5)$$

the term $l_d[\delta\psi_{n,\pm}, \delta\varphi_{n,\pm}, \alpha_{n,\pm}, \beta_{n,\pm}]$ includes contributions proportional to $\Gamma_0$, $\Gamma_1$, $\gamma_0$, $\gamma_1$ and $\gamma_2$. Since their treatment is analogous, we focus only on the $\Gamma_0$ term as an illustrative example:

$$\begin{aligned}
l_d[\delta\psi_{n,\pm}, \delta\varphi_{n,\pm}, \alpha_{n,\pm}, \beta_{n,\pm}] \supset (2S)^2\Gamma_0\Big[&\frac{1}{4}(\delta\rho_{n-1,+} + \delta\rho_{n,+} - \delta\rho_{n-1,-} - \delta\rho_{n,-})^2 + (\rho^0)^2(\alpha_{n-1,+} + \alpha_{n,+} - \alpha_{n-1,-} - \alpha_{n,-})^2 \\
&- i\rho^0(\delta\rho_{n-1,+} + \delta\rho_{n,+} + \delta\rho_{n-1,-} + \delta\rho_{n,-})(\alpha_{n-1,+} + \alpha_{n,+} - \alpha_{n-1,-} - \alpha_{n,-})\Big].
\end{aligned} \quad (6)$$

By carrying out procedures (ii) and (iii) in the main text, we obtain the full expression of the Keldysh action for fluctuating variables.

$$S = \int dtdx \mathcal{L}_1^w[\delta\rho_c, \delta\zeta_c, \alpha_q, \beta_q] + \mathcal{L}_2[\delta\rho_q, \delta\zeta_q, \alpha_c, \beta_c] \tag{7}$$

$$\begin{aligned}\mathcal{L}_1^w[\delta\rho_c, \delta\zeta_c, \alpha_q, \beta_q] &= i(\delta\rho_c \partial_t \alpha_q + \delta\zeta_c \partial_t \beta_q) + i(2S)^{\frac{3}{2}} J\kappa \delta\rho_c \nabla(\nabla\alpha_q - \beta_q) + i\frac{2SJ^2}{\Gamma_0}\delta\zeta_c(\nabla\alpha_q + \beta_q) \\ &+ 2S J^2 \kappa^2 (\frac{1}{\Gamma_0}(\nabla\alpha_q)^2 + \frac{1}{\gamma_0}(\nabla\beta_q)^2) + i\frac{t^2}{\kappa^2 \gamma_0}\delta\zeta_c \nabla^2 \beta_q + \frac{1}{2}m_\beta^2 \beta_q^2,\end{aligned} \tag{8}$$

$$\begin{aligned}\mathcal{L}_2[\delta\rho_q, \delta\zeta_q, \alpha_c, \beta_c] &= i(\delta\rho_q \partial_t \alpha_c + \delta\zeta_q \partial_t \beta_c) - i(2S)^{\frac{3}{2}} J\kappa \delta\rho_q \nabla(\nabla\alpha_c + \beta_c) + i\frac{2SJ^2}{\Gamma_0}\delta\zeta_q(\nabla\alpha_c + \beta_c) \\ &+ \frac{1}{2}(2S)^2 \Gamma_0 (\nabla\delta\rho_q)^2 + \frac{1}{2}\frac{2SJ^2}{\gamma_0}(\nabla\delta\zeta_q)^2 - i\frac{t^2}{\kappa^2 \gamma_0}\zeta_q \nabla^2 \beta_c \\ &+ \frac{1}{2}m_\sigma^2 \delta\rho_q^2 + \frac{1}{2}m_\Delta^2 \delta\zeta_q^2,\end{aligned} \tag{9}$$

where $m_\beta^2 = 8S\Gamma_1 \kappa^2$, $m_\sigma^2 = 16S^2 \gamma_1$ and $m_\Delta^2 = 16S^2 \gamma_2$ are the mass terms.

We then perform the Gaussian integral to complete step (iv), thereby obtaining Eq. (10) in the main text. This can be understood heuristically by noting that in Eq.(9), the phase couplings for $\delta\rho_q$ and $\delta\zeta_q$ share the same form. Consequently, integrating out $\beta_c$, leads to a cancellation of the leading term $\delta\rho_q \nabla^2 \alpha_c$. leaving the sub-leading term $\delta\rho_q \nabla^4 \alpha_c$. In contrast, in Eq.(8), the phase couples differently to $\delta\rho_c$ and $\delta\zeta_c$. Consequently, integrating out $\beta_c$ preserves the leading term $\delta\rho_c \nabla^2 \alpha_q$. This preservation is a direct consequence of the weak dipole symmetry, a point we will revisit when discussing the strong dipole symmetry case.

The action (10) in the main text can be directly generalized to higher dimensions for the following reasons. First, The Euler-Lagrange equation in higher dimension is virtually identical to the Eq.(2), with the index $n$ and $n+1$ simply extending to nearest neighbors in a higher-dimensional lattice. The sole remaining concern is derivative mixing across axes of different dimensions. Such mixing could only arise between the fields $\psi$ and $\varphi$; however, their coupling term contains only a single derivative, which precludes any derivative mixing. This is also consistent with the symmetry argument: to preserve the rotational symmetry near the fixed point, the relevant or marginal terms that can be constructed in higher dimensions are almost those already present in the main-text action (10).

## 2. PERTURBATIVE RG ANALYSIS

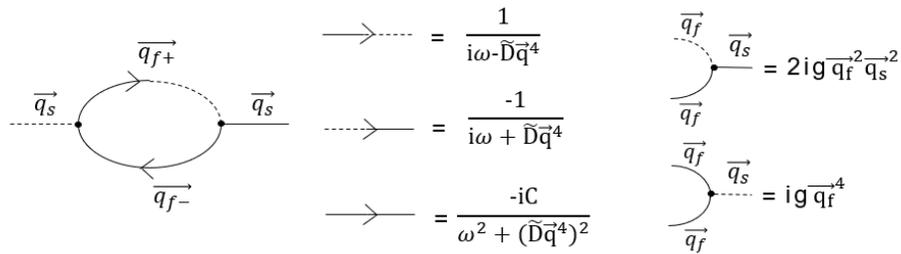

FIG. 1. One-loop Feynman diagram for the quadratic term $\delta\rho_q \nabla^2 \alpha_c$, and the Feynman rules.

In this section, we first perform a tree-level renormalization group (RG) analysis of the interacting effective action. We then demonstrate that the quadratic term $\delta\rho_q \nabla^2 \alpha_c$ is generated at the one-loop level.

We start with the interacting effective action:

$$\tilde{\mathcal{L}}_2 = i\delta\rho_q(\partial_t\alpha_c + \tilde{D}\nabla^4\alpha_c - g(\nabla^2\alpha_c)^2) + \frac{C}{2}\delta\rho_q^2. \tag{10}$$

We first perform a tree-level RG analysis near the Gaussian fixed point. Simple power counting yields the scaling dimension $[g] = \frac{4-d}{2}$, indicating that the interaction term must be retained. We therefore proceed to a one-loop analysis. We now demonstrate that the one-loop RG correction generates the quadratic term $\delta\rho_q\nabla^2\alpha_c$.

$$\begin{aligned}
\delta\tilde{D}_2 &= \frac{-8g^2}{i}\int\frac{d^dq_f d\omega}{(2\pi)^{d+1}}\vec{q}_s^{\,2}\vec{q}_{f-}^{\,4}\vec{q}_{f+}^{\,2}D_0^R(\vec{q}_{f+},\omega)D_0^K(\vec{q}_{f-},\omega) \\
&= \frac{-8g^2 C}{\tilde{D}^2}\int\frac{d^d q_f}{(2\pi)^d}\frac{\vec{q}_s^{\,2}\vec{q}_{f+}^{\,2}}{\vec{q}_{f+}^{\,4}\vec{q}_{f-}^{\,2}} \\
&= \frac{-4g^2 C}{(2\pi)^d\tilde{D}^2}\frac{S_d(\Lambda)\delta\Lambda}{\Lambda^2}\vec{q}_s^{\,2} + o(\vec{q}_s^{\,3}).
\end{aligned} \tag{11}$$

Here $D_0^R(\vec{q},\omega) = \frac{1}{i\omega - \tilde{D}\vec{q}^4}$, $D_0^K(\vec{q},\omega) = \frac{-iC}{\omega^2 + (\tilde{D}\vec{q}^4)^2}$ are retarded, Keldysh Greens function, respectively, and $S_d$ is the surface area of d-dimensional ball [2]. A non-zero $\delta\tilde{D}_2$ implies the generation of the quadratic term $\delta\rho_q\nabla^2\alpha_c$ in the low-energy effective theory. The corresponding Feynman diagram and Feynman rules are provided in FIG.1.

### 3. NUMERICAL EVIDENCE OF DETERMINISTIC EQUATION

In this section, we present stronger numerical evidence for the nonlinear deterministic equation (15) from the main text.

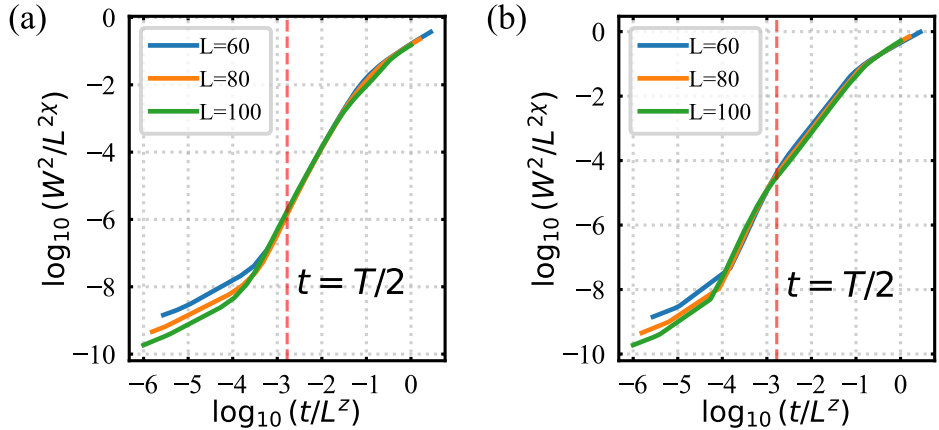

FIG. 2. Numerical simulations of the Langevin equation (12) (main text) for system sizes $L \in \{60, 80, 100\}$. The stochastic term $\xi$ is removed at time $t = T/2$, after the system has entered the scaling region. Parameter sets are: (a) $g = 0.5$, $\tilde{D}_2 = 0$, $\tilde{D} = 1$, $C = 1$; (b) $g = 2.0$, $\tilde{D}_2 = 0$, $\tilde{D} = 1$, $C = 1$. The results are plotted by fixing the scaling exponents $z = 2$ and $\chi = 2$. Together with Fig. 2(b) in the main text clearly manifest the emergence of a nonlinear deterministic equation in the long-time limit.

We remove the stochastic term at halftime during the evolution. The resulting dynamics, shown in FIG. 2(a), are essentially identical to those in FIG. 2(b) of the main text, confirming the validity of the deterministic equation. Furthermore, as shown in FIG. 2(b), similar long-time behavior is observed for a different interaction strength $g$, demonstrating the universality of this result.

In addition, we emphasize that the numerical simulations above are performed under open boundary condition, since periodic boundary condition would explicitly break the dipole symmetry. The results shown in FIG. 3 are fully consistent with this expectation.

### 4. EFFECTIVE ACTION WITH STRONG DIPOLE SYMMETRY

In this section, we examine the new constraint imposed by strong dipole symmetry and show that it results in sub-diffusive transport. We also provide details of the expansion, which conclude the effective action with strong dipole symmetry.

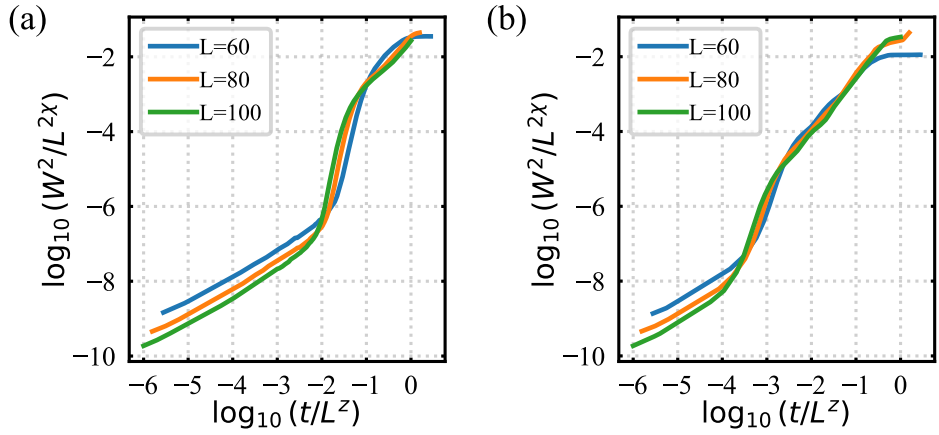

FIG. 3. Numerical simulations of the Langevin equation (12) (main text) under periodic boundary conditions for system sizes $L \in \{60, 80, 100\}$. Two parameter sets are used: (a) $g = 0.5$, $\tilde{D}_2 = 0$, $\tilde{D} = 1$, $C = 1$; and (b) $g = 2.0$, $\tilde{D}_2 = 0$, $\tilde{D} = 1$, $C = 1$. The results are plotted by fixing the exponents $z = 2$ and $\chi = 2$. The use of different $g$ values serves as an indicator of universality. The results clearly show that the system with periodic boundary conditions does not fall into the $z = 2$, $\chi = 2$ universality class, as it lacks dipole symmetry.

To understand the new constraint from strong dipole symmetry, it is useful to first recall the effective action with weak dipole symmetry Eq. (8). In that case, integrating out the field $\delta\zeta_c$ leaves the term $\delta\rho_c\nabla(\nabla\alpha_q - \beta_q)$ non-vanishing, so that $\nabla^2\alpha_q$ remains the leading term. This structure is a direct consequence of the weak dipole symmetry dissipator, which contributes a term $\delta\rho_c\nabla^2\alpha_q$ to the action. For the strong dipole symmetry dissipator, however, the same coupling form of the phase for $\delta\rho_c$ appears:

$$\mathcal{L}_1^w[\delta\rho_c, \delta\zeta_c, \alpha_q, \beta_q] = i(\delta\rho_c\partial_t\alpha_q + \delta\zeta_c\partial_t\beta_q) + ia\delta\rho_c\nabla(\nabla\alpha_q + \beta_q) + ib\delta\zeta_c(\nabla\alpha_q + \beta_q) \\ + ic\delta\zeta_c\nabla^2\beta_q + d(\nabla\alpha_q + \beta_q)^2 + e(\nabla\beta_q)^2, \tag{12}$$

where $a, b, c, d, e$ are combinations of microscopic parameters, whose explicit forms are not essential for the prediction of universal features.

After integrating out $\delta\zeta_c$ and $\beta_q$, the leading term $\delta\rho_c\nabla^2\alpha_q$ vanishes, promoting the sub-leading term $\delta\rho_c\nabla^4\alpha_q$ to dominance in the low-energy region. This, combined with the classical phase fluctuating mode, constitutes the full effective action with strong dipole symmetry (provided in the main text) and underlies the sub-diffusive behavior.

$$\mathcal{L}_1^s[\delta\rho_c, \alpha_q] = i\delta\rho_c(\partial_t\alpha_q - D\nabla^4\alpha_q) + \tilde{C}(\nabla^2\alpha_q)^2/2,$$
$$\mathcal{L}_2[\delta\rho_q, \alpha_c] = i\delta\rho_q(\partial_t\alpha_c + \tilde{D}\nabla^4\alpha_c) + C\delta\rho_q^2/2. \tag{13}$$

---

* These authors contributed equally to this work.
† These authors contributed equally to this work.
  yukezhanga08@gmail.com
‡ PengfeiZhang.physics@gmail.com


\fi
  \end{document}